\documentstyle[12pt]{article}
\def\three_j(#1,#2,#3,#4,#5,#6){\pmatrix{#1 & #2 & #3\cr
                                         #4 & #5 & #6\cr}}

\def\qqq{\end{document}}
\def\pmb#1{\setbox0=\hbox{$#1$}%
\kern-.025em\copy0\kern-\wd0
\kern.05em\copy0\kern-\wd0
\kern-.025em\raise.0433em\box0 }
\def\b{\pmb}

\def\xara(#1,#2,#3,#4){\left(\matrix{#1 & #2\cr #3 & #4\cr}\right)}
\def\Fn{J_{(n)}}
\def\Fs{J_{(s)}}
\def\Hn{H_{(n)}}
\def\Hs{H_{(s)}}
\def\thru#1{\mathrel{\mathop{#1\!\!\!/}}}
\def\CN{\cal N}
\def\w{\omega}
\def\W{\Omega}
\def\six_j(#1,#2,#3,#4,#5,#6){\left\{\matrix{#1 & #2 & #3\cr
                                         #4 & #5 & #6\cr}\right\}}
\def\W{\Omega}

\def\Ener(#1,#2){ \sqrt{{#1}^2+{#2}^2} }

\def\overlay#1#2{\setbox0=\hbox{$#1$}\setbox1=\hbox to
\wd0{\hss$#2$\hss}#1%
\hskip -1\wd0\copy1}
\def\bold#1{\setbox0=\hbox{$#1$}%
      \kern-.025em\copy0\kern-\wd0
      \kern.05em\copy0\kern-\wd0
      \kern-.025em\raise.0433em\box0 }

\def\Tr{\, \hbox{Tr} \, }

\def\bra{\langle}
\def\ket{\rangle}
\def\shalf{\,1/2\,}
\def\half{\, {1 \over 2} \,}

\def\S11{S_{11}(1535)}
\def\E0+{E_{0^+}}

\renewcommand{\thefootnote}{\fnsymbol{footnote}}
\def\footnoterule{\kern-3pt \hrule width \hsize \kern2.6pt}

\begin{document}

\title{Interference Effects in Relativistic Inclusive \\
Deuteron Electrodisintegration\thanks
{This work is supported in part by funds provided by
 the U.~S.~Department of Energy (D.O.E.)
under contract \#DE-AC02-76ER03069.}}
\author{\\ G.~I.~Poulis$^1$ and T.~W.~Donnelly$^2$ \\
\\ $^1$ {\it TRIUMF, Theory Group} \\
        {\it 4004 Wesbrook Mall} \\
        {\it Vancouver, B.C. ~ V6T 2A3, Canada} \\
\\ $^2$ {\it Center for Theoretical Physics} \\
        {\it Laboratory for Nuclear Science and Department of Physics} \\
        {\it Massachusetts Institute of Technology} \\
        {\it Cambridge, MA ~ 02139, U.S.A.}}
\date{}
\maketitle

\vfill

\centerline{Submitted to: {\it Nuclear Physics A}}
\vspace*{3cm}
\hbox to \hsize{TRI-PP-93-101
 \quad / \quad CTP\#2249 \hfil December, 1993}
\thispagestyle {empty}
\eject

\setcounter{footnote}{0}
\vspace*{.75in}
\begin{center}
{\LARGE Interference Effects in Relativistic Inclusive} \\
\vspace*{12pt}
{\LARGE Deuteron Electrodisintegration\footnotemark}

\footnotetext{This work is supported in part by funds provided by
the U.~S.~Department of Energy (D.O.E.)
under contract \#DE-AC02-76ER03069.}

\vspace*{1.0cm}
G.~I.~Poulis$^1$ and  T.~W.~Donnelly$^2$ \\

\vspace*{1.0cm}

  $^1$ {\it TRIUMF, Theory Group\\
  4004 Wesbrook Mall\\
  Vancouver, B.C. V6T 2A3, Canada}

\vspace*{0.5cm}

  $^2$ {\it Center for Theoretical Physics\\
        Laboratory for Nuclear Science and Department of Physics\\
         Massachusetts Institute of Technology\\
         Cambridge, MA 02139, U.S.A.}

\vspace*{1.0cm}
\end{center}

\vfill

\begin{abstract}
We extend the relativistic plane--wave impulse approximation formalism to
incorporate a specific class of relativistic interference effects for use
in describing inclusive electrodisintegration of $^2$H.
The role of these ``exchange'' terms for the various response functions
accessible in parity--conserving and --violating inclusive
processes is investigated and shown, especially for the latter, to have
important consequences for experiment. An extension to a simple
quasi--deuteron model is also considered.
\end{abstract}

\vfill

\thispagestyle {empty}
\eject

%%%%%%%%%%%%%%%%%%%%%%%%%%%%%%%%%%%%%%%%%%%%
\renewcommand{\thefootnote}{\arabic{footnote}}
\setcounter{footnote}{0}

\section*{1. Introduction}

One of the long--standing theoretical problems in the study
of  electron scattering processes is the incorporation of relativistic
effects  in the modeling~\cite{Gross,Tjon,BeckAren,Wallace},
in part due to our incomplete knowledge of a relativistic Hamiltonian
that realistically describes the strong interaction
dynamics~\cite{Foldy,Fearing} and in part due to the technical difficulty of
solving such problems involving the relativistic many--body problem.
On the other hand, the electromagnetic part of the interaction can
--- at least in principle --- be described in a covariant way.
It is the requirement that the level of the nonrelativistic approximations
used in the description of the strong interaction aspects of the reaction
and the one used in the electromagnetic operators be consistent
 that calls for the truncation of the nonrelativistic expansion
of the one-- and two--body electromagnetic operators
to be used between nonrelativistic nuclear states~\cite{Math,Aren79,Hadji}.
For low-- and even medium--energy studies of nuclei this has largely proven
to be a successful way to proceed.
However, there is increasing capability for electron scattering experiments
at high energies and momentum transfers where
 relying  on  leading--order nonrelativistic approximations seems
rather dangerous~\cite{Dytman,Adam,Schaar}.

While awaiting the advent  of realistic relativistic nuclear
wave functions  one might hope that
kinematical regions exist where the relativistic effects that occur when
probing a nucleus with a high momentum
transfer virtual $\gamma$ (or $Z^0$, see below) could largely ``decouple''
from the strong interaction part of the interaction, namely, from the nuclear
transition matrix elements. The quasielastic
(QE) region seems to serve that purpose: for a given momentum transfer $q$
the energy transfer $\omega$ is so chosen as to transfer the 4--momentum
essentially to one nucleon at the time, that is, along the kinematic
line where $\omega=|Q^2|/2M$ with $Q^2=\omega^2-q^2$ and $M=$ nucleon mass.
 As a result, the roles of meson--exchange currents (MEC) and
 final--state interactions (FSI) are minimized~\cite{Aren79,Hadji,Schaar}.
Upon making two extra assumptions, namely that of a plane--wave
final state and of incoherently summing  over all single--nucleon
contributions, it can be shown that the cross section
factorizes into a product of a single--nucleon half--off--shell
cross section and a spectral function, the latter
basically describing the probability of finding a nucleon
with a given energy and momentum in the target nucleus~\cite{Diep,Frull}.
This is the factorized plane--wave impulse approximation (PWIA). The
relativistic
corrections enter this model in a very different way for the spectral
function and the single--nucleon cross section. In particular,
whereas the scale that governs the relativistic effects in the
former is basically dictated by the ratio of the Fermi momentum $p_F$ to
the nucleon mass, namely $\eta_F\equiv p_F/M$ which ranges from about 0.06
in the deuteron to 0.28 in heavy nuclei, the latter
directly contains the effects of the high momentum transfer and so involves the
ratio $\kappa\equiv q/2M$ which can be large --- indeed ``high momentum
transfer'' may be defined as $\kappa \sim$ unity or larger. To the extent that
this model is roughly correct, the important
relativistic effects associated with high momentum transfer can be addressed
by treating the single--nucleon part relativistically while relying on
traditional nonrelativistic descriptions of the spectral function.
In a series of recent papers~\cite{Hadji,Caballero,Caballero_2}
we have pursued these ideas. In particular, the validity of the PWIA
was tested by comparing it with a nonrelativistic calculation.
When supplemented by earlier studies~\cite{Aren79} that verify the
suppression of the role of MEC in the QE region, these comparisons have
led us to conclude that the model is suitable for
describing high momentum transfer quasifree processes for parity--conserving
electrodisintegration, if not for parity--violating electrodisintegration
under all circumstances (see Ref.~\cite{Hadji} and below).

On the other hand, when these ideas are carried to non--quasifree
kinematics, in comparisons between the PWIA and
more complete calculations the observed discrepancies are often
interpreted in terms of the lack of FSI and MEC effects in the former.
However, it is also relevant to ask how much of this discrepancy can in
fact be attributed to the other assumption of the PWIA model, namely that of
incoherence. One argument given is that for quasifree kinematics
the interference effects should be suppressed by an extra power of
 $Q^2$~\cite{Frull,JA}. Whereas this renders the contribution of
such interference effects insignificant for deep--inelastic scattering,
it still
leaves the question open for scattering in the vicinity of the QE peak
at high but not asymptotic momentum transfers,
where it may still be important and in fact compete with
the relativistic corrections we may wish to incorporate by means of the PWIA.
Another argument invoked is that the interference effects in coincidence
processes involve the amplitude where the virtual photon interacts with
the $A-1$ residual nuclear state and would therefore be suppressed
due to the rapidly falling form factor of this state~\cite{Frull} if
this daughter system ``sticks together''.
While this may be true for heavy nuclei at low missing energy, it obviously
does not apply to few--body systems and it likely does not apply as well
to heavy nuclei at high missing energy.

The purpose of this paper is to investigate the significance of these
``exchange'' effects for the deuteron in a simple, but relativistic model
which is an extension of the PWIA in a way that includes such interference
terms.  Although our goal is to apply
this formalism to inclusive scattering, we first obtain the coincidence
response functions and then integrate over the detected nucleon's quantum
 numbers to obtain the inclusive answer. In so doing, we will recover the
plane--wave ``Born'' approximation (PWBA) of Fabian and
Arenh\"ovel~\cite{Aren79,Aren82}, although our answer will be an extension
of that earlier work as we include terms in all orders in $1/M$
for the single--nucleon electromagnetic current. We then calculate the
inclusive responses and provide a simple interpretation for the significance
of exchange effects in inclusive electrodisintegration of deuterium.
We especially wish to clarify the roles of the kinematics and the
form factors in suppressing or enhancing the importance of the
exchange terms and to stress the important part
that these terms play in forward--angle, parity--violating
 electron scattering. It is in fact this observation,
combined with the attention drawn recently to high momentum transfer
PV electron scattering experiments from few--body systems
aimed at measurements of the  strangeness form factors~\cite{Cebaf_d}, that
provides one of the main motivations for this work.

Finally, we consider a simple extension of our approach to include a
treatment of a relativistic version of the quasi--deuteron model.
Our interest in the present work
where this model is concerned is rather focused: we explore the nature of
the interference terms and relativistic effects in a situation where
$\eta_F$, being connected as it is to the characteristic Fermi momentum
for a many--body nucleus, is much larger than it is for the deuteron.

We have organized this paper as follows: in Sec.~2, we review the
basic formalism for electron scattering in the PWIA and then in Sec.~3
we modify one of the assumptions that lead to the PWIA and
 obtain a variation of it which, following Arenh\"ovel~\cite{Aren82},
we call ``the relativistic plane--wave Born approximation'' (relativistic
PWBA). In Sec.~4 we
investigate the role of the interference contributions for
different response functions in inclusive parity--conserving and
--violating electron scattering; additionally, the nonrelativistic limit
of the PWBA is obtained in order for the physics to become transparent.
In Sec.~5 the interplay between the interference and relativistic
effects is explored for a wide range of kinematics and in
Sec.~6 we briefly apply our formalism to the quasi--deuteron model.
Finally, in Sec.~7 we present our summary and conclusions.

%%%%%%%%%%%%%%%%%%%%%%%%%%%%%%%%%%%%%%%%

\section*{2. The Plane--Wave Impulse Approximation (PWIA)}

Let us begin with a brief review of the coincidence electron scattering
formalism within the context of the PWIA for a general nucleus; later we
specialize to the case of deuterium. More detailed derivations and discussions
can be found in the literature~\cite{Caballero,Rask_Don,Don_Rask,y_scaling}.
The kinematics for electrodisintegration
processes are presented in Fig.~1.
Here an electron with four--momentum $K^{\mu}\equiv(\epsilon,{\pmb k})$
is scattered through an angle $\theta_e$ to four--momentum
$K'^{\mu}\equiv(\epsilon',{\pmb k}')$. We restrict ourselves to one--photon
exchange and plane--wave electrons. For the hadronic
variables we have: $P_i^{\mu}\equiv(E_i,{\pmb 0})$ and
$P_B^{\mu}\equiv(E_B,{\pmb p_B})$. Both energies are ``on--shell'', that is
$E_i=M_i$ and $E_B=\sqrt{ \pmb p_B^2 + M_B^2}$; moreover,
the outgoing on--shell nucleon has $P_N^{\mu}=(E_N, \pmb p_N)$
with $E_N=\sqrt{\b p_N^2 + M^2}$.
The four--momentum transfer is $Q^{\mu}=(\w,{\pmb q})$ where
$\w=\epsilon-\epsilon'$ and ${\pmb q} = {\pmb k}-{\pmb k}'$. We will
denote the magnitude of the momentum transfer by $q=|{\pmb q}|$.

The cross section for exclusive electron scattering in the laboratory
system can be written as~\cite{Rask_Don}
\begin{equation}\label{excl_cross}
{d\sigma\over d\Omega_e d\epsilon' d\Omega_N} = {2 \alpha^2\over Q^4}
\left(\epsilon'\over \epsilon\right){p_N M M_B\over (2\pi)^3 M_i}
f^{-1}_{rec}\eta_{\mu\nu} W^{\mu\nu} .
\end{equation}
Here   $f_{rec}=1+{2\epsilon\over M_i}\sin^2{\theta_e\over 2}$
 is  the hadronic recoil factor and
 $\eta_{\mu\nu}$ is the leptonic tensor
\begin{equation}\label{lepton}
\eta_{\mu\nu}= K_{\mu}K'_{\nu} + K'_{\mu}K_{\nu} + {1\over 2} Q^2
g_{\mu\nu} -ih \epsilon_{\mu\nu\alpha\beta}K^{\alpha}K'^{\beta} ,
\end{equation}
with $h=\pm1$ the electron helicity (only longitudinally polarized
electrons enter in the relativistic limit $ {m_e/\epsilon}\rightarrow 0$;
see Ref.~\cite{Don_Rask}).
The hadronic tensor $W^{\mu\nu}$ contains all information about
 the nuclear structure and
dynamics and is constructed from the electromagnetic (EM) current
transition matrix elements as
\begin{equation}\label{hadron}
W^{\mu\nu}=(J^{\mu}_{fi})^*(J^{\nu}_{fi})=
{\overline{\sum}}_{i,f}\bra f|\hat J^{\mu}|i\ket^*
\;\bra f|\hat J^{\nu}|i\ket \ ,
\end{equation}
where $|i\ket$ represents the target state and $|f\ket$ the final state
containing the residual nucleus and the emitted nucleon.
The contraction of the two tensors
can be written as
\begin{equation}\label{contra}
\eta_{\mu\nu} W^{\mu\nu}= {1\over 2} v_0 [ R_{fi} + h R'_{fi}]\ ,
\end{equation}
where $v_0=4\epsilon\epsilon' \cos^2{\theta_e\over 2}$.
The quantities $R_{fi}$ ($R'_{fi}$) contain the parts of the tensors in
Eqs.~(\ref{lepton}) and (\ref{hadron}) that are, respectively,
 symmetric and antisymmetric under the
interchange of $\mu$ and $\nu$. It can be shown that they can be decomposed
in a sum of response functions multiplied by leptonic kinematical
factors~\cite{Don_Rask}:

\begin{eqnarray} \label{responses}
R_{fi} & = & v_L R^L_{fi} + v_T R^T_{fi} + v_{TL} R^{TL}_{fi}
        + v_{TT} R^{TT}_{fi} \\
R'_{fi} & = &  v_{T'} R^{T'}_{fi} + v_{TL'} R^{TL'}_{fi} \ .
\end{eqnarray}
Different responses contribute to different processes, depending on the
exclusivity and the presence/absence of polarization degrees of freedom;
in this work we will not be especially concerned with polarized targets.
In particular, for the case of inclusive, unpolarized target,
parity--conserving scattering only the transverse ($T$) and
longitudinal ($L$) response functions  contribute, while for inclusive
parity--violating scattering (necessitating a polarized electron beam)
one also gains access to a $T'$ response~\cite{Hadji,Walecka}.
The $TL$, $TT$ and $TL'$ responses enter in coincidence processes and
when hadronic polarizations are involved~\cite{Rask_Don,Don_Rask}.
Explicit formulae in the general case for the kinematical factors $v_i$ and
the definition of the responses $R^i_{fi}$  in terms of components of the
hadronic tensor can be found in those references.

One is now confronted with the problem of calculating the
hadronic tensor in Eq.~(\ref{hadron}). As stated in the Introduction,
in this work we wish to extend the PWIA in a direction that incorporates
specific interference effects. We therefore begin by reviewing the
approximations that lead to the factorized PWIA. In so--doing we draw upon the
discussion in Ref.~\cite{Caballero}. Specifically, the PWIA contains three
approximations: one--body current operators, plane--wave final state
(hence, no FSI), and
the assumption that the detected nucleon is the one that reacted with the
virtual photon~\cite{deForest}.
Under the first assumption one has for the electromagnetic current
 operators~\cite{Frull}
\begin{equation}\label{1_body}
\hat J_{\mu} = \sum_{m,m'} \sum_{\tau,\tau'} \int d{\pmb u}
\int d{\pmb u}' \bra{\pmb u}', m',\tau'|\hat j_{\mu}| {\pmb u}, m, \tau\ket
a^{\dagger}_{ {\pmb u}' m' \tau'} a_{ {\pmb u} m \tau} \ ,
\end{equation}
where $|{\pmb u},m,\tau\ket$ is an on--shell nucleon spinor
with momentum $\pmb u$
and spin projection $m=\pm\half$ referring to an axis of quantization along
$\pmb q$, unless otherwise specified.
The label $\tau=\pm\half$ denotes the isospin of the nucleon.
 The one--body operator $\hat j_{\mu}$ represents the physics of
the $\gamma N N$ vertex and reads $\hat j_{\mu} = \exp{(iQ_\alpha X^\alpha)}
\Gamma_{\mu}(Q^2)$. The vertex function  $\Gamma_{\mu}(Q^2)$
contains the Dirac structure of the struck nucleon.

The second and third assumptions can be summarized in the formula
\begin{equation}\label{PW_and_inco}
\bra f| a^{\dagger}_{ {\pmb p}' m' \tau'} =\delta({\pmb p_N} - {\pmb p}')
\delta_{m_N,m'}\delta_{\tau_N,\tau'}\bra B|
\end{equation}
with the following interpretation: the final state
 $|f\ket$ contains a plane--wave emitted nucleon plus the residual state
$|B\ket$, supplemented with  the {\it additional assumption}
 that the emitted nucleon
is the nucleon to which the virtual photon is connected.
It is precisely this last assumption that we wish to relax in
order to obtain the PWBA. Before doing so let us
briefly see how the above three assumptions lead to the familiar factorized
PWIA~\cite{Caballero,deForest}. Inserting
Eq.~(\ref{1_body}) and using Eq.~(\ref{PW_and_inco}) to calculate
$\bra f|\hat J_{\mu}|i\ket$, we find
\begin{equation}\label{matrix_elementa}
\bra f|\hat J_{\mu}|i\ket = \sum_m \int d{\pmb u}
\bra {\pmb p_N}, m_N| \hat j^{\mu}_{\tau_N}| {\pmb u}, m\ket
\bra B| a_{ {\pmb u} m\tau_N}| i\ket\ .
\end{equation}
Upon squaring to obtain the hadronic tensor we find
\begin{equation}\label{PWIA_te}
W^{\mu\nu}({\pmb p_N},{\pmb q}) =
 \sum_{m,m'} {\cal W}^{\mu\nu}_{m m'}({\pmb p},{\pmb p_N})\;
 \eta_{m m'}({\pmb p})
\delta^3({\pmb p} +{\pmb q}-{\pmb p_N}) \ ,
\end{equation}
where we define the {\it single--nucleon tensor}
\begin{equation}\label{S_N_T}
{\cal W}^{\mu\nu}_{m m'}({\pmb p},{\pmb p_N}) = \sum_{m_N}
\bra {\pmb p_N}, m_N|\Gamma^{\mu}| {\pmb p}, m'\ket^*
\bra {\pmb p_N}, m_N|\Gamma^{\nu}| {\pmb p}, m \ket
\end{equation}
and the {\it  momentum distribution}
\begin{equation}\label{spectral}
\eta_{m m'}({\pmb p}) = \sum_B \bra B| a_{{\pmb p} m' }|i\ket^*
 \bra B| a_{{\pmb p} m }|i\ket \ ,
\end{equation}
where the spin content has been explicitly
taken into account. It can be shown that when the target is
unpolarized both the single--nucleon tensor and the momentum distribution
are diagonal in the spin sector, that is $m=m'$~\cite{Caballero,deForest}.
Under these assumptions the electromagnetic interaction
occurs only via the single--nucleon tensor which,
apart from  uncertainties due to the off--shellness of the
struck nucleon $|{\pmb p},m\ket$~\cite{Naus},
 can be computed relativistically, as stated in the Introduction.
A graphical depiction of the PWIA approximation is presented in Fig.~2a.

%%%%%%%%%%%%%%%%%%%%%%%%%%%%%%%%%%%%%%%%

\section*{3. The Plane--Wave Born Approximation (PWBA)}

We now relax the assumption that the detected nucleon is the same as the
one that was struck by the virtual photon and restrict
 ourselves to a deuteron target.
 The physical interpretation of relaxing the above assumption
in the case of the deuteron can be seen in Fig.~2. Specifically,
suppose for the moment that one performs a coincidence $^2$H$(e,e'$p)n
measurement. While a proton is known to be detected, it might happen that it
was either the proton (Fig.~2a) or the neutron (Fig.~2b) to which the
photon connected. Some aspects of this possibility have been taken into
account in the past (see for example \cite{Frull} and references
therein\footnote{In a recent preprint Hummel and Tjon\cite{Hummel}
have also gone further in addressing the issue of relativistic effects in
parity--conserving deuteron electrodisintegration within a quasipotential
framework where exchange (Born) terms and FSI are taken into account; in
our work we focus on the former for both parity--conserving and
parity--violating
inclusive electrodisintegration, but do not consider the latter.  The
similarity of the results obtained when the same responses are considered
reinforces our emphasis on the incorporation of the exchange effects via
the relativistic PWBA. Note that the (more limited) approach taken in the
present work can also be applied straightforwardly to nuclei with
$A>2$~\cite{Adam}.}),
although one does not usually take these exchange effects into account at the
same time treating the single--nucleon vertex  to all orders in $1/M$. Indeed,
the calculations of Arenh\"ovel~\cite{Aren82}  and Fabian and
Arenh\"ovel~\cite{Aren79} are the closest
to our approach in that they retain terms up to order $1/M$
in the electromagnetic operators;
here our goal is to incorporate specific relativistic
ingredients to all orders while exploring such exchange effects and hence
this aspect of our work is a natural extension of those previous studies.
In addition, while these ideas could
be applied to coincidence electron scattering, our present focus is on
the roles that relativity and exchange play for inclusive processes (which
are less commonly addressed in terms of exchange effects) and,
consequently, after this section we shall specialize to the reactions
$^2$H($e,e^\prime$)pn and $^2$H(${\vec e},e^\prime$)pn, the latter
involving parity--violating electron scattering.

We proceed by replacing Eq.~(\ref{PW_and_inco}) by
\begin{eqnarray}\label{PW_coh}
\bra f|a^{\dagger}_{{\pmb p}',m',\tau'} &=&
\delta^3({\pmb p}'-{\pmb p_N})
\delta_{m',m_N}\delta_{\tau',\tau_B} \bra {\pmb p_B}, m_B,\tau_B|\nonumber\\
& &\mbox{} -\; \delta^3({\pmb p}'-{\pmb p_B})
\delta_{m',m_B}\delta_{\tau',\tau_B} \bra {\pmb p_M}, m_M,\tau_M| \ .
\end{eqnarray}
Here the undetected nucleon in the deuteron is labelled ``B'' in order to
connect with the conventions of Sec.~2. The minus sign in front of the
second term is there to ensure antisymmetrization of the two--fermion state.
Using the one--body {\it ansatz}, Eq.~(\ref{1_body}),
we obtain for the electromagnetic matrix elements
\begin{eqnarray}\label{PWBA_ME}
\bra f| \hat J_{\mu}| d\ket &=&\sum_m\sum_{\tau}\Biggl\{
 \int  d{\pmb u}\, \bra {\pmb p_N}, m_N,\tau_N |\hat j_{\mu}|{\pmb u},
m,\tau\ket
\bra {\pmb p_B},m_B,\tau_B|\,a_{{\pmb u},m,\tau}\,|d\ket\nonumber\\
& & \mbox{} -\int \, d{\pmb u} \bra {\pmb p_B}, m_B,\tau_B|
\hat j_{\mu}|{\pmb u}, m,\tau\ket
\bra {\pmb p_N},m_N,\tau_N|\,a_{{\pmb u}, m,\tau}| d\ket\Biggr\} \ .
\end{eqnarray}
We now calculate the break--up amplitudes for a deuteron
with the following quantum numbers: isospin
$T=0$ and $M_T=0$, orbital angular momentum $L\in\{0,2\}$ coupled with
spin $S=1$
to total angular momentum $J=1$ and prepared in a magnetic substate $M_J$:
\begin{eqnarray}\label{break_up}
\lefteqn{
\bra {\pmb p_1},m_1,\tau_1|\,a_{{\pmb p_2},m_2,\tau_2}\,|J,M_J\ket  = }\\
& &\qquad{\CN}\;\delta^3({\pmb p_1} +{ \pmb p_2})\sum_{L=0,2}\;\sum_{M_L=
-L}^{+L}
i^L \,u_L(p_1) Y^{M_L}_L(\W_{\pmb p_1})(-1)^{1+M_J+M_S}\times\nonumber\\
& &\sum_{M_S=-S}^{+S} \three_j(L,S,J,M_L,M_S,-M_J)
\three_j(\half,\half,S,m_1,m_2,-M_S)\three_j(\half,\half,T,\tau_1,\tau_2,-M_T)
\ .\nonumber \end{eqnarray}
Here $\CN$ is an overall normalization factor
and $u_L(k)$ the Bessel transforms of the
S-- and D--state radial wave functions of the deuteron
(for $L=0$ and $L=2$ respectively). Inserting into Eq.~(\ref{PWBA_ME})
we obtain (using the fact that the single--nucleon matrix elements
in Eq.~(\ref{PWBA_ME}) are diagonal in isospin space)
\begin{equation}\label{matrix_element}
\bra f|J^{\mu}|d\ket =
\sum_m \biggl\{   A_m^B\bra {\pmb p_N}, m_N| \Gamma_{\tau_N}^{\mu}|
{\pmb p}, m\ket
 +  A_m^N\bra -{\pmb p}, m_B| \Gamma_{\tau_B}^{\mu}|
-{\pmb p_N}, m\ket\biggr\} \ ,
\end{equation}
where we define for $i\in\{N,B\}$
\begin{eqnarray}\label{alfas}
A_m^i &\equiv & {{\cal N}\over \sqrt{2}}(-1)^{1/2-\tau_N}
\delta_{\tau_N,-\tau_B}
\sum_L \sum_{M_L}\sum_{M_S}\three_j(L,1,1,M_L,M_S,-M_J)
\times\nonumber \\
& &\;(-1)^{1+M_J+M_S} u_L(p_i) Y_L^{M_L}(\W_{\pmb p_i})
\three_j(\shalf,\shalf,1,m_i,m,-M_S)\ .
\end{eqnarray}
In the following we will freely use ${\pmb p_B}+{\pmb p}=0$. The
physics content of Eq.~(\ref{matrix_element}) can be made more transparent
by examining Fig.~2.

The  hadronic tensor for the unpolarized coincidence process where
the nucleon $i=N$ is detected then reads
(after we sum over the final--state quantum numbers and average
over the initial--state ones):
\begin{equation}\label{direx}
W^{\mu\nu}={1\over {2J+1}}\sum_{M_J}\sum_{m_N}\sum_{\tau_B}\sum_{m_B}
\sum_m\sum_{m'}\biggl\{
D^{\mu\nu} + D'^{\mu\nu} + E^{\mu\nu} + E'^{\mu\nu} \biggr\} \ ,
\end{equation}
where we have defined two ``Direct'' and two ``Exchange'' tensors as
\begin{eqnarray}\label{def_tensors}
        D^{\mu\nu}& \equiv & A^{B*}_{m'}\; A^B_m\;
\bra {\pmb p_N}, m_N| \Gamma_{\tau_N}^{\mu}|-{\pmb p_B}, m'\ket^*\;
\bra {\pmb p_N}, m_N| \Gamma_{\tau_N}^{\nu}|-{\pmb p_B}, m\ket\\
        D'^{\mu\nu}& \equiv & A^{N*}_{m'} \;A^N_m\;
\bra {\pmb p_B}, m_B| \Gamma_{\tau_B}^{\mu}|-{\pmb p_N}, m'\ket^*\;
\bra {\pmb p_B}, m_B| \Gamma_{\tau_B}^{\nu}|-{\pmb p_N}, m\ket\nonumber\\
        E^{\mu\nu}& \equiv & A^{B*}_{m'}\; A^N_m\;
\bra {\pmb p_N}, m_N| \Gamma_{\tau_N}^{\mu}|-{\pmb p_B}, m'\ket^*\;
\bra {\pmb p_B}, m_B| \Gamma_{\tau_B}^{\nu}|-{\pmb p_N}, m\ket\nonumber\\
        E'^{\mu\nu}& \equiv & A^{N*}_{m'}\; A^B_m\;
\bra {\pmb p_B} ,m_B| \Gamma_{\tau_B}^{\mu}|-{\pmb p_N}, m'\ket^*\;
\bra {\pmb p_N}, m_N| \Gamma_{\tau_N}^{\nu}|-{\pmb p_B}, m\ket\ .\nonumber
\end{eqnarray}
 A diagrammatical interpretation of these tensors is given in Fig.~3.
The direct contributions amount to incoherent scattering,
whereas the two exchange contributions incorporate
the interference terms. It is important to notice
that  the {\it coincidence} PWIA
corresponds to just one of the direct terms, namely $D^{\mu\nu}$.  We first
obtain
\begin{eqnarray}\label{alfas_K}
        \lefteqn{
\sum_{\tau_B}\sum_{M_J}A_{m'}^{j*} A_m^i ={1\over{2}}{\cal N}^2
\sum_{L,L'}\sum_{M_L,M_L'}\sum_{M_S,M_S'} \;\sum_{K=0,1,2}\;
\sum_{X=-K}^{+K} (2K+1) (-1)^{M_L+M_S+1}}\nonumber\\
& &\times\three_j(1,1,K,M_S,-M_S',X)u_L(p_i) u_{L'}(p_j)\six_j(L,L',K,1,1,1)
Y_L^{M_L}(\W_{\pmb p_i}) Y_{L'}^{*M_L'}(\W_{\pmb p_j})\nonumber\\
& &\times\three_j(L,L',K,M_L,-M_L',-X)\three_j(1/2,1/2,1,m_i,m,-M_S)
\three_j(1/2,1/2,1,m_j,m',-M_S')\ .
\end{eqnarray}
Let us  calculate the direct terms ($i=j$). It can be shown that
only $K=0$ contributes in this case. This also forces diagonality in
spin space, $m=m'$,
and one is left with the expression
\begin{eqnarray}\label{PWIA_tensor}
\sum_{spins}D^{\mu\nu}&= & {{\cal N}^2\over {24\pi (2J+1) }}
\sum_L |u_L(p)|^2 \sum_{m_N}
\sum_{m,m'}\delta_{m,m'}\\
& & \times\Tr\biggl\{ |{\pmb p},m\ket\bra {\pmb p},m'|\gamma^0
\Gamma^{\mu\dagger}_{\tau_N}
\gamma^0|{\pmb p_N},m_N\ket\bra{\pmb p_N},m_N|\Gamma^{\nu}_{\tau_N}\biggr\}
\nonumber \ ,
\end{eqnarray}
for the PWIA term and a similar one for the other direct term $D'^{\mu\nu}$
which, however, does not contribute to the coincidence PWIA, since if $D$
is being calculated for $(e,e'p)$ then $D'$ corresponds to $(e,e'n)$
and {\it vice versa.\/}
Note that the spin sums decouple between the single--nucleon
matrix elements and the break--up amplitude $3-j$ symbols. The
PWIA answer for the single--nucleon tensor corresponding to
the reaction $^2$H$(e,e'$N)B is then simply
\begin{equation}\label{PWIA_final}
{\cal W}^{\mu\nu}_{PWIA} = {{\cal N}^2\over {24\pi (2J+1) (2 M)^2}}
\Tr\biggl\{ [{\thru p}+M] \gamma^0\Gamma^{\mu\dagger}_{\tau_N}
\gamma^0 [{\thru {p}_N+M}]\Gamma^{\nu}_{\tau_N} \biggr\} \ ,
\end{equation}
where comparison with Eq.~(\ref{PWIA_te}) shows that the momentum
distribution in
this case is simply $\eta(p) \sim \sum_{L} u_L(p)^2$.

The situation is more complicated in the case of the
exchange terms for two reasons. Firstly, note  from Eq.~(\ref{def_tensors})
that both $m_N$ and $m_B$ appear in the single--nucleon parts;
moreover, the arguments in the spherical harmonics are not identical
any more. As a result, the spin sums cannot be done without the
explicit spin--dependence of the single--nucleon matrix elements.
In general, all $K$--values contribute and $m,m'$ need not be equal
as before.
Secondly, the spinor--product sums one encounters are not any longer diagonal
in momentum space (for example, one needs
$|{\pmb p_B},m\ket\bra {\pmb p_N}, m'|$). In order to overcome the
non--diagonality in the spin sector we could proceed as in
Ref.~\cite{Caballero}
where we introduced a diagonal hadronic tensor with spinors quantized
with respect to a generic direction. However, to take care of
the non--diagonality
in momentum space we would have to boost the $|{\pmb p_B},m\ket$ state
into, say, one with momentum ${\pmb p_N}$. Together with the $\gamma_5$'s
that are parts of the spin--projectors needed
(due to the decoupling of the spin sums)
to convert the single--nucleon part into a trace,
 this would lead to a trace with a large number of
$\gamma$ matrices. For these reasons we have
abandoned the idea of a trace formulation and, instead, we evaluate the
spin sums at the outset. To that end we will use the notation
of Alberico {\it et~al.} in Ref.~\cite{Alberico} and decompose the
single--nucleon operators into spin--flip ($s$) and non--spin--flip ($n$)
parts:
\begin{eqnarray}\label{torino_1}
J^{\mu}_{m_N,m_i}\equiv\bra{\pmb p_N},m_N|\Gamma^{\mu}_{\tau_N}|-
{\pmb p_B},m_i\ket \!&\!=\!&\!
\Fn^{\mu} \delta_{m_N,m_i} +\Fs^{\mu} (2 m_i) (1-\delta_{m_N,m_i})\nonumber\\
H^{\mu}_{m_B,m_i}\equiv\bra{\pmb p_B},m_B|\Gamma^{\mu}_{\tau_B}|
-{\pmb p_N},m_i\ket \!&\!=\!&\!
\Hn^{\mu} \delta_{m_B,m_i} +\Hs^{\mu} (2 m_i) (1-\delta_{m_B,m_i})
 \ .\nonumber\\
\end{eqnarray}
For the 3--vector currents in the above equation we choose a coordinate system
with the 3--axis along the momentum transfer ${\pmb q}$, the 2--axis along
${\pmb v} ={\pmb q}\times {\pmb p}$
(that is, perpendicular to the hadron plane)
and the 1--axis along ${\pmb v}\times {\pmb q}$ (in the hadron plane).
\begin{equation}\label{3_currents}
{\pmb J}_{m_j,m_i} =
J^{(1)}_{m_j,m_i} \hat{\pmb u}_1  +
(2 m_i)J^{(2)}_{m_j,m_i} \hat{\pmb u}_2  +
J^{(3)}_{m_j,m_i} \hat{\pmb u}_3 \ ,
\end{equation}
and a similar equation for ${\pmb H}_{m_j,m_i}$. Detailed  expressions for
the matrix elements in Eq.~(\ref{torino_1}) in terms of
the single--nucleon form factors can be found in the Appendix.
We can now perform all the spin sums for both the direct-- and
exchange--type terms. The sums in the former case are trivial
(as expected) and one ends up with the following formula for the
direct term, which should be
compared with Eq.~(\ref{PWIA_tensor}):
\begin{equation}\label{direct_sum}
\sum_{spins}D^{\mu\nu}=\cases{
0\;,{\hskip 1.5 true in} \hbox{if ${\mu}=2$ or ${\nu}=2$}&, but not both \cr
\noalign{\medskip}
{{\cal N}^2\over {72 \pi}}\sum_L |u_L(p_B)|^2
\Bigl\{ \Fn^{*\mu }\Fn^{\nu } + \Fs^{* \mu }\Fs^{\nu}\Bigr\} &,
otherwise \cr}
\end{equation}
and a similar expression for the other direct term
\begin{equation}\label{direct_p_sum}
\sum_{spins}D'^{\mu\nu}=\cases{
0\;,{\hskip 1.5 true in} \hbox{if ${\mu}=2$ or ${\nu}=2$}&, but not both \cr
\noalign{\medskip}
{{\cal N}^2\over {72 \pi}}\sum_L |u_L(p_N)|^2
\Bigl\{ \Hn^{*\mu }\Hn^{\nu} + \Hs^{*\mu}\Hs^{\nu}\Bigr\} &,
otherwise.\cr}
\end{equation}
We finally proceed to obtain the two exchange terms.
After evaluating the spin summations we can cast both exchange contributions
in the compact form
\begin{equation}\label{tildes}
    \sum_{spins}[E^{\mu\nu}+E'^{\mu\nu}] = {\cal E}^{\mu\nu}
   +{\cal E}^{*\nu\mu} \ ,
\end{equation}
where we have set
\begin{eqnarray}\label{exchange_sum}
{\cal E}^{\mu\nu}
&=&{1\over 6}{\cal N}^2\sum_{K}\sum_{X}\sum_{L,L'}
   u_L(p) u_{L'}(p_N) \sum_{M_L,M_L'}\six_j(L,L',K,1,1,1)
(2K+1) (-1)^{M_L}
\nonumber\\
& &\times Y_L^{M_L}(\W_{\pmb p})
Y_{L'}^{*M_L'}(\W_{\pmb p_N})\three_j(L,L',K,M_L,-M_L',-X) \nonumber\\
&  &\times \sum_{i,j\in\{n,s\}} a^{K,X}_{i,j} H^{*\mu }_{(i)} J^{\nu}_{(j)} \ .
\end{eqnarray}
The coefficients $a^{K,X}_{i,j}$ are defined in the Appendix and
appear in Table~I.

%%%%%%%%%%%%%%%%%%%%%%%%%%%%%%%%%%%%%%%%

\section*{4. The S--State--Only, Nonrelativistic, Inclusive PWBA}

Eq.~(\ref{exchange_sum}) cannot be further simplified
unless certain approximations are made. In the next--to--leading order
 nonrelativistic limit
it reproduces the results of Arenh\"ovel (Eqs.~(A.1)--(A.5) in
Ref.~\cite{Aren82}), notwithstanding the fact that our calculation is
performed in the laboratory frame whereas the one in Ref.~\cite{Aren82}
is performed in the center--of--mass frame.  Our aim here is to gain some
understanding of the difference between the PWBA and PWIA
in the {\it inclusive} case. The inclusive PWIA is defined to be the sum
of the  individual contributions obtained by
integrating over final momenta in the $^2$H(e,e$^\prime$p)n and
$^2$H(e,e$^\prime$n)p
cross sections. In other words, the inclusive PWIA amounts to the
{\it total incoherent} part of the PWBA, unlike the coincidence situation,
where the $D'^{\mu\nu}$ term in Eq.~(\ref{def_tensors}),
 although belonging to the incoherent
contribution, is {\it not} part of the coincidence PWIA.

We first wish to obtain a qualitative understanding of the role
of interference terms in the inclusive  electrodisintegration of
deuterium. To that end here we shall make several approximations; of course,
these approximations are not made when presenting results in later sections.
 Referring to Eq.~(\ref{exchange_sum}) and Table~I we first note that
restricting ourselves to the S--state--only yields considerable
simplification as then $K=X=0$ and also all
angular dependence becomes trivial. It should be noted that this
is not an unreasonable
approximation for {\it quasielastic}, inclusive
 electron scattering. The reason is that
 the strength of the inclusive QE response comes predominantly from
 the low momentum components of the struck nucleon~\cite{y_scaling},
whereas the D--state contribution (in momentum space)
 becomes comparable to the S--state--only
around 250 MeV/c, more than 4 times the deuteron Fermi momentum
of about 55 MeV/c (which
is manifested in the width of the QE response~\cite{Fermi_Gas}).
The form of the PWBA longitudinal and transverse responses in the
S--state--only limit is given in the Appendix.
We now make a further simplification in this initial look at the
responses and restrict ourselves to the
leading nonrelativistic contributions for $J^{\mu}_{(i)}$, $H^{\mu}_{(i)}$,
which as seen in Table~II involve the electric and magnetic electromagnetic
single--nucleon
form factors $G_{E,M}^\tau$, where $\tau=\tau_N$ or $\tau_B$ labels
the isospin projection (see Ref.~\cite{Alberico}).
The ``3'' components are eliminated by current
conservation~\cite{Don_Rask} and we define $\kappa\equiv q/2M$.

In the coordinate system we have chosen, the longitudinal and transverse
 responses are given in terms of the following components of the hadronic
tensor~\cite{Caballero}: $R_L = W^{00}$ and $R_T = W^{11} + W^{22}$.
Referring now to Eqs.~(\ref{longitudinal_s_only},\ref{transverse_s_only})
in the Appendix we can write for the longitudinal response in the
nonrelativistic, S--state--only limit:
\begin{equation}\label{long_1}
R_L= { {\cal N}^2\over {72 \pi}}\biggl\{ u^2(p_B){G_E^{\tau_N}}^2
 + u^2(p_N){G_E^{\tau_B}}^2 + 2 u(p_N)u(p_B) G_E^{\tau_N} G_E^{\tau_B}
\biggr\} \ .
\end{equation}
Similarly, for the transverse response we have
\begin{equation}\label{tran_1}
R_T= { {\cal N}^2\over {72 \pi}} \times \kappa^2
\biggl\{ u^2(p_B){G_M^{\tau_N}}^2
 + u^2(p_N){G_M^{\tau_B}}^2 + \left(2\over 3\right)
 u(p_N)u(p_B) G_M^{\tau_N} G_M^{\tau_B}
\biggr\} \ .
\end{equation}
The same nonrelativistic limits have been obtained
 with a simple calculation in Refs.~\cite{Hadji,Thesis}.
In order to obtain the inclusive EM responses, we
integrate over final momenta including the energy--conserving delta function
$\delta(\w-\W)$ where
\begin{equation}\label{wmega}
\W({\pmb p_N,\pmb p_B})=\Ener(\pmb p_N,M)+\Ener(\pmb p_B,M)-M_d =
\W({\pmb p_B,\pmb p_N})\ .
\end{equation}
We define the following ``Direct'' and ``Exchange'' integrals
${\cal D}$ and $\cal E$
\begin{eqnarray}\label{integrals}
{\cal D} &=& \int d{\pmb p} \delta(\w-\W) u^2(p) \nonumber\\
{\cal D'} &=& \int d{\pmb p} \delta(\w-\W) u^2(p_N)= {\cal D}\nonumber\\
{\cal E} &=& \int d{\pmb p} \delta(\w-\W) u(p)u(p_N) \ ,
\end{eqnarray}
where as always ${\pmb p_B} + {\pmb p_N}=0$ and Eq.~(\ref{wmega})
guarantees that ${\cal D}={\cal D'}$.
Then for the inclusive responses we have
\begin{eqnarray}\label{inclu}
R_L&\sim &\;\biggl\{ {G_E^{T=0}}^2\Bigl({\cal D} +{\cal E}\Bigr)
+  {G_E^{T=1}}^2\Bigl({\cal D} -{\cal E}\Bigr)\biggr\}\nonumber\\
R_T&\sim & \kappa^2\, \biggl\{ {G_M^{T=0}}^2\Bigl({\cal D}
+{{\cal E}\over 3}\Bigr)
+  {G_M^{T=1}}^2\Bigl({\cal D} -{{\cal E}\over 3}\Bigr)\biggr\}\ ,
\end{eqnarray}
where we have omitted an overall constant. The factor of 3 in the
exchange part of the transverse responses is a consequence of the
spin--flip character of the relevant nonrelativistic
operators~\cite{Thesis} and the fact that the spin of the deuteron is 1.
 The importance of this factor in suppressing the
role of the exchange terms has been also observed by Frankfurt and Strikman
in a light cone calculation~\cite{Frank}.
Comparing the above expressions with the inclusive PWIA, we note that the
 latter is defined as the  sum of the integrals over final momenta of
the $^2$H(e,e$^\prime$p)n and $^2$H(e,e$^\prime$n)p
cross sections, and therefore reproduces
just  the  ${\cal D}$  terms of Eq.~(\ref{inclu}).

\subsection*{4.1. The $\cal E/D$ Ratio}
This suggests that the ${\cal E}/{\cal D}$ ratio
sets the na{\"\i}ve scale for the importance of the interference terms.
We say na{\"\i}ve, because the actual scale for the various
response functions can be very different from  ${\cal E}/{\cal D}$,
due to the specifics
of the form factors involved, as we shall discuss shortly. To begin with
we examine the behaviour  of this ratio as a function of the
kinematical variables $q$ and $\w$. Let us  first perform the angular
integrations in Eq.~(\ref{integrals}) using the energy--conserving
delta function,
 and rewrite the Direct and Exchange integrals in the form
\begin{eqnarray}\label{integrals_2}
{\cal D}&=&2\pi\int_{|y(q,\w)|}^{Y(q,\w)} p dp {E_N(p)\over q} u^2(p)
\nonumber \\
{\cal E}&=&2\pi\int_{|y(q,\w)|}^{Y(q,\w)}
 p dp {E_N(p)\over q} u(p) u(p_N)
 \ ,
\end{eqnarray}
where
\begin{equation}\label{basikh}
E_N(p)=\sqrt{p_N^2+M^2} = M_d + \w -\sqrt{p^2  + M^2} \ .
\end{equation}
The transition from Eqs.~(\ref{integrals}) to Eqs.~(\ref{integrals_2}),
 as well as the limits of
integration are discussed in Refs.~\cite{y_scaling,Thesis}.
There are two questions we wish to address:
firstly, what is the behaviour of ${\cal E/D}$ with $\w$ for fixed
momentum transfer $q$, and secondly,
what is the behaviour of ${\cal E/ D}$ as a function
of the  momentum transfer $q$ for fixed QE kinematics, $\w\approx |Q^2|/2M$?
 Strictly speaking, the
range of $\w$ values for which we can legitimately address the first
question in the context of a plane--wave, no MEC model, is restricted
to the regime of quasifree scattering as characterized by the width of
the QE peak,
$\Delta\w=\sqrt{2}q p_F/ \sqrt{M^2+q^2}$~\cite{Hadji}.

We first plot in Fig.~4 the integrands of the Direct and Exchange
contributions for a small value of momentum transfer, $q=150$ MeV/c.
Here we plot $pu^2(p)$
and $p u(p) u(p_N)$, which are, respectively, the integrands of
${\cal D}$ and $\cal E$
 in the nonrelativistic limit (up to a constant), where $u(p)$ is
the Bessel transform of the S--state deuteron wave function.
It is clear that, being linear in $p$, the direct integrand
starts from zero at the origin,
then peaks due to the fact that $u(p)$ drops dramatically with $p$,
and eventually dies off with larger values of $p$.
On the other hand, the form of the exchange integrand, ${\cal E}$,
 for fixed momentum transfer $q$,  depends on the value of
 the energy transfer $\w$,
since for each given momentum $p$ we must solve Eq.~(\ref{basikh})
 in order to find the corresponding value of $p_N$. We show
this integrand for
two characteristic values of $\w$, namely, one close to  threshold, where
$\w_{thr}+M_d= \sqrt{ (A M)^2 + q^2}$
with $A=2$ for the deuteron, and one on the quasielastic peak, where
$\w_{qe} + M_d= \sqrt{ M^2 + q^2} + M$~\cite{y_scaling}.
To understand the  form of the integrands, we need to know approximately
where $p_N(q,\w,p)=p$, since this will give us the point where the
${\cal E}$ and ${\cal D}$ integrands intersect.
In this particular case $ M\gg q$, since $q=150$ MeV/c
and $M\approx 939$ MeV,
and ignoring the small $(\approx 2.25$ MeV)
 deuteron binding energy, we can solve Eq.~(\ref{basikh})
at threshold as
\begin{equation}\label{THR}
         p^2 + p_N^2 = {q^2\over 2}\ ,
\end{equation}
that is, $p=p_N $ at $p=q/2$,
which means that the two integrands cross around $p=q/2=75$ MeV/c,
as indeed is seen in the Fig.~4a.
 On the other hand, the point where $p=p_N$ in the case where $\w=
\w_{qe}$
is found from Eq.~(\ref{basikh}) as
\begin{equation}\label{QE}
         p^2 + p_N^2 = {q^2}\ ,
\end{equation}
that is, $ p=p_N$ at $p=q/ \sqrt{2}$,
which is approximately at $p=107$ MeV/c, as seen in the Fig.~4b.
As far as the limits of integration are concerned,
in the case of the deuteron they are given by~\cite{Thesis}
\begin{eqnarray}\label{limits}
p_{min}=|y|,\quad& &\hbox{where}\; y=(M_d+\w)\sqrt{{1\over 4}
        -{M^2\over W^2}}-{q\over 2}\nonumber\\
p_{max}= Y, \quad& &\hbox{where}\; Y=y+q \ ,
\end{eqnarray}
with  $W^2 = (M_d+\w)^2 - q^2$ being the center--of--mass energy.
On the peak, $y=0$~ and therefore the limits extend from
$0$ to $q$~\cite{y_scaling}.
 Since the two integrands cross at $107$ MeV/c (a high value
compared to the Fermi momentum $p_F = 55$ MeV/c,
which characterizes the falloff of $u(p)$)
and both start from zero, we learn two things:
first, that  ${\cal D}$ will be a maximum on the quasielastic peak,
(hence the very notion of the {\it peak} for $\w =\w_{qe}$),
and second that the ${\cal E}/{\cal D}$
ratio on the peak will be rather small. This is illustrated in Fig.~4b,
 where we have graphically depicted the two integrals by
shading the corresponding integrands.

However the situation is entirely different near threshold.
There, the limits of integration are $p_{min}=p_{max}=q/2$, which means
that as we approach threshold from above, the two integrals vanish;
what is interesting, however, is that their {\it ratio} approaches 1.
This happens because it is precisely at $p=q/2$ that the two
integrands are equal, as we argued above. That means that for $\w$
close to threshold, we will be integrating in an area just around
$p=q/2$, where by continuity the two integrands will be very similar and
therefore the ${\cal E}/{\cal D}$ ratio will be close to 1.
We depict this situation in Fig.~4a
for $\w=9$ MeV, where the limits are $p_{min}= 47$ MeV/c and
$p_{max}= 102$ MeV/c. We can clearly see that the two shaded areas,
corresponding to the two integrals, are very similar.

We summarize this behaviour of the   ${\cal E/ D}$  ratio
as a function of energy transfer, for  two cases $q=150$ MeV/c
and $q=300$ MeV/c in Figs.~5a and 5b.
In order to provide a feeling for the individual integrals, and not just their
ratio, we have included Tables~III.1 and III.2.
 The position of the QE peak for those $q$ values
 is at 14.13 and 49.89 MeV, respectively.

Let us next examine the behaviour of the ${\cal E/ D}$ ratio
 as a function of the momentum transfer $q$, on the quasielastic peak.
 It is precisely around the quasielastic peak that we hope to be able to
trust a quasifree calculation like the plane--wave
approximation, and an understanding of the role of the
interference terms as given by the Exchange integral
 in that kinematical region therefore seems necessary.
We begin by noticing that the larger the value of momentum transfer,
the further apart are $p$ and $p_N(p,\w_{qe}(q))$ located.
To see why this happens, let us recall that in terms of the
$y$--scaling variable
we have~\cite{Hadji}
\begin{equation}\label{y_var}
M_d+\omega= E_B + E_N= \sqrt{M^2 + (q+y)^2} +
\sqrt{M^2 + y^2} \  .
\end{equation}
Examining the direct term  we see that the integral
has its maximum when $y=0$ and thus $\omega=\omega_{qe}$.
For fixed $y=0$ the answer is independent of $q$, since the
only other $q$--dependence in this simplified model is contained in $Y$ and
becomes negligible when $q$ and $Y\rightarrow\infty$. On the other hand,
the exchange term involves an overlap
between $u(p)$ and $u(p_N)$. At $y=0$ we have $E_B +E_N = \sqrt{M^2+q^2}
+M$, and therefore, as $q$ becomes large so does the sum of $ E_B =
\sqrt{p^2+M^2}$ and $E_N = \sqrt{{p_N}^2+M^2}$.  Hence, not both of $p$ and
$p_N$ can be small at the
 same time.  Since the function $u(p)$ is
localized to very small values of $p$ ({\it i.e.,\/} most of the momentum
distribution lies at $p < p_F\sim$ 55 MeV/c), this
immediately implies that the ratio ${\cal E}/{\cal D}$ must fall with
increasing $q$.
This is to be expected, since the importance of the interference terms
%is in some sense a measure of the ``classicality'' of the process, which
scales roughly like $q/p_F$.
The ratio ${\cal E}/{\cal D}$, which sets the na{\"\i}ve scale for
the relative importance of the exchange terms, is shown in Fig.~6
for kinematics corresponding to the impulse approximation position of
the QE peak,
\begin{equation}\label{peak}
\omega_{qe} = \sqrt{M^2 + q^2} + \epsilon_d - M  \ ,
\end{equation}
where $\epsilon_d = 2M-M_d$ is the deuteron binding energy
The characteristic
momentum where this ratio becomes 1/2 is about 110 MeV/c $\sim 2 p_F$.
This is suggestive of a simple Fermi gas picture where
two momentum distributions with characteristic width $p_F$ overlap
considerably only when they are centered at momenta differing
by less than $2 p_F$.
We see that the ratio drops dramatically with increasing
momentum transfer and thus we should expect that the exchange contributions
for deuterium become unimportant beyond, say, 400--500 MeV/c and accordingly
that the assumption of ignoring the interference terms of the PWBA
will become reasonable at sufficiently high $q$. However, at these high
momentum transfer values we need to include the effects of the D--state
and relativity we have ignored so far. This we will discuss in Sec.~5.

\subsection*{4.2. The role of form factors}

Our previous analysis suggests that, at least around the quasielastic peak
and at relatively
high momentum transfer values, the differences between the de--relativized
PWIA and PWBA should disappear.
However,  caution should be exercised~\cite{Hadji,Thesis} even when studying
only the region near the QE peak, since the significance of the ratio
${\cal E}/{\cal D}$ may be quite different
for the various response functions and, ultimately, for the
parity--conserving cross section and parity--violating asymmetry, depending
on the choice of kinematics.

Consider first the longitudinal EM response:
for small momentum transfers, $ G_E^{(T=0)}\simeq G_E^{(T=1)}$, since $G_{En}$
is very small. Now we see from Eq.~(\ref{inclu}) that
it is $\left(G_E^{(T=0)^2}+G_E^{(T=1)^2}\right)$
 that multiplies  ${\cal D}$,
and $\left(G_E^{(T=0)^2}-G_E^{(T=1)^2}\right)$ that multiplies ${\cal E}$.
As we have seen,  ${\cal E/D}$ is appreciable at
small $q$ values, precisely where the form factor
combination that multiplies it is negligible,
and therefore $R_L^{(T=0)} + R_L^{(T=1)}$ is basically the
same whether the PWBA or PWIA is used.
 When $G_{En}$ becomes appreciable, ${\cal E}/{\cal D}$ has already
dropped dramatically, and thus the conclusion still holds.
On the other hand, a different argument is found
for the transverse EM terms, where $G_M^{(T=0)}$ and
$G_M^{(T=1)}$ are rather different (equivalently, $G_{Mp}$ and $G_{Mn}$ are
both large and different).  As we can see from Eqs.~(\ref{inclu}),
there the scale is given by ${\cal E}/3{\cal D}$, which suppresses the
exchange effects, and
hence we  see that the PWBA and PWIA results should not be
much different as long as ${\cal E}/{\cal D}$ is not too large.  Similar
conclusions hold for the transverse PV responses which are also
spin--flip--dominated.

However, the results obtained
for the {\it electroweak} response functions that appear in the
numerator of the parity--violating asymmetry~\cite{Hadji}
in the process $^2$H(${\vec e},e'$)np
may be very different. In particular, the electroweak longitudinal
response $\tilde R^L $, whose contribution in the PV asymmetry
is most important at forward angles,
can be very sensitive to ${\cal E}/{\cal D}$ effects. This is due to the
particular form of the weak neutral current couplings
$\beta_V^{(T=0,1)}$ in terms of which the electroweak form factors become
\begin{eqnarray}\label{betas}
\tilde G^{T=0}_{E,M} &=&\beta_V^{(0)} G^{T=0}_{E,M}+
        \beta_V^{(s)} G^{(s)}_{E,M}\nonumber\\
\tilde G^{T=1}_{E,M} &=&\beta_V^{(1)} G^{T=0}_{E,M}\ ,
\end{eqnarray}
where $\beta_V^{(1)}=1-2\sin^2\theta_W$,
 $\beta_V^{(0)}=-2\sin^2\theta_W$ and $\beta_V^{(s)}=1$~\cite{Revart}.
Ignoring the strangeness contributions
(labelled ``(s)'') we see from Eq.~(\ref{inclu})
that the scale dictating the importance of the interference terms
for the PV longitudinal response
is changed from the na{\"\i}ve value ${\cal E}/{\cal D}$ to
\begin{equation}\label{telos}
\left({\beta_V^{(0)} - \beta_V^{(1)}\over \beta_V^{(0)} +
\beta_V^{(1)}}\right)
{{\cal E}\over {\cal D}} = {-1\over 1-4\sin^2\theta_W}
{{\cal E}\over {\cal D}} \ ,
\end{equation}
which, due to the  rather interesting coincidence
 that $\sin^2\theta_W\sim 0.227$ ({\it i.e.,\/} nearly 1/4 where the factor
$1-4\sin^2 \theta_W$ goes to zero)
 is about $-11{\cal E}/{\cal D}$ in the standard model~\cite{Hadji}.
For this reason,
 the parity--violating longitudinal response
can be extremely different in the PWIA and PWBA,
 even leading to opposite
signs for the asymmetry obtained near threshold using the two models,
as we have demonstrated in Ref.~\cite{Hadji}.
This is illustrated in Fig.~7 where we plot
the PV asymmetry for forward scattering ($\theta_e=35^{\circ}$)
at $q=150$ and 300 MeV/c using  both the PWIA and PWBA models:
whereas the results for the cross section
are indistinguishable (not shown), the PV asymmetries
are extremely different for the two approximations, especially as one goes
towards threshold.  For more discussion of the PV asymmetry, see
Refs.~\cite{Hadji,Revart}.

%%%%%%%%%%%%%%%%%%%%%%%%%%%%%%%%%%%%%%%%

\section*{5. The Role of Relativity}
\bigskip
In this section we present results for kinematics not restricted to the
quasielastic region. For such kinematics important corrections to the
plane--wave approximations such as final--state interactions, meson exchange
currents and pion production effects can be important.  In the present
work our focus is not on such aspects of the full problem, but rather
on a specific model where {\it both relativistic and exchange\/} effects
can be incorporated at the same time. To that end we define the following
ratios for subsequent use, involving the fully--relativistic Born ($BF$)
and Impulse ($IF$) models, as well as
a fully--relativistic S--state--only version ($BSF$ and $ISF$) and the
nonrelativistic S--state--only Born and Impulse reductions
 discussed in the previous section ($BN$ and $IN$, respectively):
\begin{equation}\label{del_ratio}
 \Delta_{BF/IF}\equiv \left | {BF/IF}-1 \right | \ ,
\end{equation}
which is a measure of the difference between the Born and Impulse
approximation, and
\begin{equation}\label{r_ratio}
 \Delta_{BSF/BN}\equiv \left |{BSF/BN-1}\right | \ ,
\end{equation}
which is a measure of the importance of the relativistic corrections.

We start our presentation of the full results with the absolute value of
the ratio $\cal{E}/\cal{D}$ which
sets the na{\"\i}ve scale for the importance of exchange effects. In Fig.~8
we plot percentage contours for this ratio, for $100\leq q\leq 800$ MeV/c
and energy transfer $\w$ from threshold
up to $max\{q,400$ MeV$\}$. In all of the following contour plots we
indicate the position of the quasielastic ridge.
 One observes that the
ratio has a trough at low $q$ which coincides with the
quasielastic ridge and a clear ridge near threshold,
 as anticipated from the discussion
in the previous section and Fig.~4. Thus, the $\cal{E}/\cal{D}$
ratio for quasifree kinematics is less than 10\% for $q$ above
250--300 MeV/c. Moving away from the quasifree region we observe
 another ridge for large values of energy transfer near the real--photon line.
This maximum is due to the specifics of the deuteron wave functions used
(Reid soft core) and is not reproduced by simplistic wave functions (see
next section).

Next we discuss the $\Delta_{BF/IF}$ ratio for the two responses accessible in
unpolarized EM inclusive scattering, $R_L$ and $R_T$. In the following we
use the Galster parametrization for the nucleon form factors~\cite{Galster}.
The $R_L$ results are presented in Fig.~9. It is clear that unless one stays
 very close to threshold, the
longitudinal response is almost totally oblivious to the exchange terms.
The only remnant of the second ridge in the $\cal{E}/\cal{D}$ ratio in
Fig.~8 is a 1\% effect, which, however, lies outside of the quasifree region.
Thus our qualitative analysis of Sec.~4 tracing the insensitivity
of $R_L$ to the exchange effects back to the fact that  $G_{En}$ is negligible
at those values of $Q^2$ where  $\cal{E}/\cal{D}$ is appreciable proves
to be adequate even for the full calculation. This at first
sight seems strange, since the relativistic single--nucleon
matrix elements now also have $G_M$ contributions. Referring to the
Appendix, however, we see that the dominant contribution of that type will be
$G_{Ep} G_{Mn} \sinh\Phi \sin\Psi \cosh\Phi \cos\Psi$, which is
proportional to $(p/M)^2$ and hence very much suppressed. For example,
with a Fermi momentum distribution $\theta(p_F-p)$, for $q>p_F$
the contribution of the $G_E$ terms to the $\Delta_{BF/IF}$ ratio for
quasielastic kinematics is proportional to $1-\left(a/p_F\right)^2$,
 with $a=\sqrt{q^2-p_F^2}$,
 whereas the contribution of the $G_M$ terms would be proportional to
 $1-\left(a/p_F\right)^4$.

 In Fig.~10 we present the
results for $R_T$. Again we observe that, except for very low values of
momentum transfer the quasielastic ridge is less than 2\% sensitive to
exchange effects. However, this changes when one moves away from
quasifree kinematics, especially towards threshold, and above
the quasielastic peak around $q=400$ MeV/c, where there is a 10\%
peak; the exchange terms there are definitely not negligible compared to
other effects such as FSI\footnote{See also the footnote at the beginning
of Sec.~3 where we point to the recent preprint of Hummel and
Tjon\cite{Hummel}.} or
MEC that are usually pointed to as shortcomings of the PWIA. We see that
the magnitude of the effect is  not the same as the $\cal{E}/\cal{D}$ ratio,
the reason being that --- unlike the longitudinal, where the suppression of
exchange effects comes from the form factors --- the leading interference
contributions are suppressed by a factor of 3 compared to the direct ones
(see Eq.~(\ref{inclu})).

We now shift our attention to the importance of the relativistic corrections
that we have included in the single--nucleon matrix elements.
Looking at the explicit expressions of these matrix elements in
the Appendix, we see that as a general rule such
relativistic corrections occur in two classes:
the first class involves terms like  $(1+\tau)$ (the Darwin
term) or like $q^2/|Q^2|$. The second class involves
terms that are proportional to $\sinh\Phi$ or $\sin\Psi$ and therefore
proportional to
$(p\sin\theta/M)$. These last terms are suppressed for two reasons:
firstly, the Fermi momentum provides a cutoff for $p$ which is especially low
in the case of the deuteron; secondly, on the
quasielastic peak we have $\w=2M\tau+2M-M_d$, and therefore
$p\sin\theta/M=\sqrt{
{|Q^2|\over q^2}({M_d+\w\over 2M})^2-(1+\tau)}$ vanishes.
Accordingly, we expect
the contribution of second--class terms to be small
in the quasifree region where the dominant contribution to the momentum
integrals comes from $p\sin\theta$ values close to 0 (this corresponds to
the common --- yet quite misleading --- statement that quasifree kinematics
amounts to the struck nucleon being at rest). However, the first--class
terms can be very important with high values of momentum transfer.
For example at the quasielastic peak corresponding to $q=1$ GeV/c
the factor $q^2/|Q^2|$ induces a $20\%$ effect.
Thus we see that there is indeed a simple reason why in the quasielastic
region the two scales that drive the relativistic corrections to the problem,
namely $q/M$ and $p/M$, decouple: the latter is always associated with a
$\sin\theta$ factor which vanishes on the quasielastic peak and is in
general small around it.
Note that, if we simply take $1/M$ as the ``one'' scale that characterizes
the relativistic effects without being careful about the momentum that
multiplies it, we would argue that it is the first--class
terms that are of order $O[1/M]^2$ and thus have to be dismissed, while
the second--class terms, being of order $O[1/M]$ have to be retained.
For example, the calculation of Ref.~\cite{Aren82} includes corrections
 up to order $[1/M]$ and thus ignores the first--class corrections while
retaining terms linear in $(p\sin\theta/M)$. This is of course {\it consistent}
with the treatment of the deuteron wave function in that work.
{\it Numerically,\/} however, we wish to point out that the
incorporation of the first--class--type corrections can be important.
The above remarks support the comments made in the Introduction where
we argue that in the quasifree region the ``dynamical'' relativistic
corrections associated with the momentum of the struck nucleon
decouple from the more ``kinematical'' factors associated with the
large momentum transfer, like $q^2/|Q^2|$. Note that were we to perform our
 calculation in the center--of--mass frame we would lose the
advantage of having the Fermi momentum cutoff as the only scale
regarding relativistic corrections for the deuteron wave function
and the struck nucleon,
as now a $q$--dependent operator would be required in order to boost
to the center--of--mass frame. Although the relativistic corrections
induced by
this boost can be shown in general not to be as significant as those
induced by using relativistic single--nucleon current
operators~\cite{BeckAren}, we prefer to minimize them by working
directly in the deuteron rest frame.

In Fig.~11a we show contour plots for the $\Delta_{BSF/BN}$ ratio (as defined
above in Eq.~(\ref{r_ratio}))
in the case of the longitudinal response. Effects of order 10\% are present
in the quasifree region for $q$ beyond 600 MeV/c, whereas for higher momentum
transfer values there can be 10\% effects already at 400 MeV/c and
20\% effects at $q=$ 700 MeV/c. We observe a ridge parallel to the
quasielastic region but at momentum transfer values greater by about
150 MeV/c. This ridge is directly related to the $G_M$ terms in
the longitudinal EM response. To see that, first recall from Fig.~9 that for
this response the exchange effects are less than 1\% (except near threshold)
 and thus decouple from the relativistic effects, allowing the
ratio $\Delta_{BSF/BN}$ to be analyzed in terms of the
PWIA instead of the PWBA. From the Appendix (see also
Refs.~\cite{Hadji,deForest})
we have for the single--nucleon longitudinal response (on--shell)
             \begin{equation}\label{R_L}
R_L^{[T=0,1]} \simeq (1+\lambda)^2{ {G_E^T}^2 + \tau {G_M^T}^2
 \over  (1+\tau)} - \kappa ^2 {G_M^T}^2  \ ,
                   \end{equation}
where $\lambda = \w/2M$ and $\kappa = q/2M$, as above. For the
deuteron response we have to sum over the $T=0$ and $T=1$ channels and
 integrate over momentum.
Defining $\xi = \left ( {G_M^{T=0}}^2+{G_M^{T=1}}^2 \over
{G_E^{T=0}}^2 + {G_E^{T=1}}^2 \right )$, which is a large number (of
order 10), we have for the $\Delta_{BSF/BN}$ ratio in the PWIA
             \begin{equation}\label{r_dyo}
\Delta_{BSF/BN}  = | {(1+\lambda)^2\over(1+\tau)} (1 + \tau \xi) -
\kappa ^2 \xi - 1 |  \ .
\end{equation}
Observe that on the quasielastic peak we have $\lambda\rightarrow \tau$
and $\kappa ^2 \rightarrow \tau(1+\tau)$ and so the $G_M$ terms cancel and
$\Delta_{BSF/BN} \rightarrow \tau$, as reflected in Fig.~11a. Away from
the quasifree region,
however, despite seemingly being suppressed by an extra factor
of $\tau$, these terms are important since the actual scale is not $\tau$ but
$\xi \tau$.
With the presence of these terms $\Delta_{BSF/BN}$ can have a ridge (it could
not have one
without these terms, since $(1+\lambda)^2/(1+\tau)$ is monotonically
increasing with $\w$ as $q$ is kept fixed) which is approximately at
             \begin{equation}\label{r_tria}
\lambda \simeq {1\over \xi} + \kappa ^2 \Rightarrow\w\simeq
\w _{QE} + {2M\over \xi} \ ,
                   \end{equation}
with $2m\xi\cong 100$ MeV/c, which describes rather well the ridge in Fig.~11a.
The importance of the large magnetic moment in counterbalancing the
small $\tau$ has been already stressed in the
past~\cite{Alberico}. We wish to draw the attention of the reader to
another point. If we include the $(1+\tau)$ factor (the Darwin term) only,
the relativistic longitudinal response is {\it smaller}
 than the nonrelativistic one~\cite {BeckAren}.
However, what we see from our expressions is quite different, as
$(1+\lambda)^2\simeq\left(E_N + E_B\over 2 M\right)^2$ is in fact
even more important, since it is equal to $(1+\tau)^2$ on the quasielastic peak
and thus our ``relativistic'' longitudinal response
is larger than the nonrelativistic one.
In fact, by {\it multiplying} the nonrelativistic response, Eq.~(\ref{inclu})
by $(1+\tau)$ and then forming a new $\Delta_{BSF/BN}$ ratio we see from
Fig.~11b that
we can include relativistic corrections in the quasifree region
very effectively as now $\Delta_{BSF/BN}$ is of order 1\% in this region.

Next we discuss relativistic corrections to the transverse response.
Contours for the ratio $\Delta_{BSF/BN}$ for the transverse response are
plotted in
Fig.~12a. The effects are important even in the quasielastic region. One
notices that the relativistic corrections have only a small $q$ dependence.
Let us attempt to understand this by using the PWIA formula for the
transverse response
          \begin{equation}\label{R_T}
R_T^{[T=0,1]} \simeq \Bigl[ {\kappa ^2\over \tau}(1+\lambda)^2 -(1+\tau)\Bigr]
 { {G_E^T}^2 + \tau {G_M^T}^2  \over  (1+\tau)} - 2 \tau {G_M^T}^2
 \ .                \end{equation}
Here, the exchange effects are not negligible and
accordingly an analysis of the
$\Delta_{BSF/BN}$ ratio using the Impulse approximation in the place of the
PWBA will only be a rough
approximation, although this still accounts for the basic trend of the
graph.  Working as before we can cast the ratio $\Delta_{BSF/BN}$ in the form
          \begin{equation}\label{T_dyo}
\Delta_{BSF/BN}  = |{- \lambda ^2\over \kappa} +
{(1+\tau\xi)\over 2\kappa ^2\xi (1+\tau)}
\Bigl[ {\kappa ^2\over\tau} (1+\lambda)^2 -(1+\tau)\Bigr] | \ .
                \end{equation}
The term in brackets is $(p\sin\theta/M)^2$ and vanishes on the quasielastic
peak. The first term, $\lambda ^2/\kappa$ simply reflects the fact that
the extreme nonrelativistic expressions in Eq.~(\ref{inclu})
have a $q^2$ instead of a $|Q^2|$ that appears multiplying the leading--order
magnetic terms in Eq.~(\ref{R_T}). This is an example of the first class
of relativistic corrections mentioned before. It results in contours parallel
to the $q$--axis. The other term in the $\Delta_{BSF/BN}$ ratio is a
second--class
correction and hence is minimal on the quasielastic peak.
Thus, by simply rewriting the nonrelativistic PWIA using $\tau$
instead of $\kappa^2$, the resulting $\Delta_{BSF/BN}$ ratio, plotted in
Fig.~12b is now
less than 1\% in the quasielastic region.

%%%%%%%%%%%%%%%%%%%%%%%%%%%%%%%%%%%%%%%%

\section*{6. The Quasi--Deuteron Case}

{}From the previous sections we have seen that
 interference effects can be suppressed because of
(a) the small Fermi momentum of the deuteron which characterizes the
${\cal E}/{\cal D}$ ratio, (b) the vanishing of $G_{En}$ at small values of
$|Q^2|$ which minimizes exchange effects in the EM
longitudinal response, or (c) the factor of $1/3$ in Eq.~(\ref{inclu})
which suppresses the transverse EM response. We have seen that
in the case of parity--violating electron scattering (b) does not
apply and the longitudinal response has a large exchange contribution.
In this section we wish to investigate another situation where one can get
sizable interference effects. Consider the quasi--deuteron model~\cite{
Heidmann,Levinger,Watson,Theis,Schoch}
 where the scattering process
can be thought as occurring off a $NN$ quasiparticle embedded in a complex
nucleus. In such a situation the spatial extent of the quasiparticle can be
much smaller than a free deuteron, resulting in a broader momentum
distribution,
which in turn results in an ${\cal E}/{\cal D}$ ratio  that  tends to drop
less rapidly with momentum transfer. Moreover, this quasiparticle can
be in a $S=0,\enskip T=1,\enskip J=0$ state, namely only an $L=0$, S--state
where the spin sums are much easier to evaluate. We obtain for the coincidence
hadronic tensor:
         \begin{equation}\label{quasi_D}
D^{\mu\nu} \sim u(p_B)^2\Bigl\{ J_{(s)}^{*\mu}J_{(s)}^{\nu}
         +J_{(s)}^{*\mu}J_{(s)}^{\nu}
\Bigr\}\Bigl[ 1 - \delta_{\mu,2} - \delta_{\nu,2} +
2\delta_{\mu,2} \delta_{\nu,2} \Bigr] \ ,
                   \end{equation}
and similarly
         \begin{equation}\label{quasi_Dp}
D'^{\mu\nu} \sim u(p_N)^2\Bigl\{ H_{(n)}^{*\mu}H_{(n)}^{\nu}
 + H_{(s)}^{*\mu}H_{(s)}^{\nu}
\Bigr\}\Bigl[ 1 - \delta_{\mu,2} - \delta_{\nu,2} +
2\delta_{\mu,2} \delta_{\nu,2} \Bigr] \ .
                   \end{equation}
Thus, there is no change in the direct terms (Cf. Eq.~(\ref{direct_sum})).
For the exchange terms one again obtains (Eq.~\ref{tildes}) with
         \begin{eqnarray}\label{quasi_E}
{\cal E}^{\mu\nu} & \sim & u(p_N)u(p_B)\Bigl\{ J_{(n)}^{*\mu}H_{(n)}^{\nu}
\Bigl[1 -\delta_{\mu,2}-\delta_{\nu,2}\Bigr] \nonumber \\
 & & - J_{(s)}^{*\mu}H_{(s)}^{\nu}
\Bigl[ 1 - \delta_{\mu,2} - \delta_{\nu,2} +
 2\delta_{\mu,2} \delta_{\nu,2} \Bigr] \Bigr\} \ .
                   \end{eqnarray}
We can now compute the longitudinal and transverse inclusive
responses: after taking the nonrelativistic limit we obtain
 the analog of Eq.~(\ref{inclu}).
         \begin{eqnarray}\label{inclu_2}
R_L&\sim &\;\biggl\{ {G_E^{T=0}}^2\Bigl({\cal D} +{\cal E}\Bigr)
+  {G_E^{T=1}}^2\Bigl({\cal D} -{\cal E}\Bigr)\biggr\}\nonumber\\
R_T&\sim & \kappa^2\, \biggl\{ {G_M^{T=0}}^2\Bigl({\cal D} -{\cal E}\Bigr)
+  {G_M^{T=1}}^2\Bigl({\cal D} +{\cal E}\Bigr)\biggr\}\ .
                   \end{eqnarray}
That is, the extreme nonrelativistic limit of the longitudinal response
remains functionally the same (the ${\cal D}$ and ${\cal E}$ integrals
change due to the different momentum distribution), but the transverse
response exchange terms are different and there is no suppression factor
of $1/3$ as in Eq.~(\ref{inclu}) ---
this is due to the fact that for a spin--0
target the same number of diagrams contributes to both the direct
 and the exchange terms. These points are made clear in Figs.~13--15
where we show the analogs of Figs.~8--10 using Eq.~(\ref{quasi_E})
and a quasi--deuteron square--well momentum distribution with Fermi momentum
300 MeV/c and well--depth 50 MeV. The ${\cal E}/{\cal D}$ ratio remains
significant up to rather high momentum transfer values in the quasifree
region, dropping to less than 20\% only after 500 MeV/c, whereas for the
usual $^3S_1$ deuteron discussed above the ratio drops below 20\% already
at about 200 MeV (inside the quasielastic ridge). The longitudinal response
does not show any significant change since the neutron form factor
suppressing effect is still present. However, the transverse response (Fig.~15)
shows a dramatic change compared to (Fig.~10) corresponding to
the deuteron, reflecting in a linear fashion the
${\cal E}/{\cal D}$ ratio (this is because ${G_M^{T=0}}^2\ll {G_M^{T=1}}^2$).

\bigskip
%%%%%%%%%%%%%%%%%%%%%%%%%%%%%%%%%%%%%%%%

\section*{7. Summary and Conclusions}

In this work we have relaxed one of the assumptions inherent
in the plane--wave impulse approximation to electron--nucleus scattering,
namely, the identification of the detected nucleon with the nucleon
struck by the virtual photon, while retaining specific aspects of
relativity. This we call the relativistic plane--wave Born approximation.
For inclusive  electron scattering from deuterium
(where we integrate over the detected nucleon's quantum numbers)
the difference between the PWBA and the PWIA amounts to the interference
between the amplitudes corresponding to the case where the struck and
detected nucleons  are the same and the one where they are not.
We calculate the hadronic tensor in the PWBA treating the single--nucleon
matrix elements relativistically and the deuteron wave function
nonrelativistically. The relativistic effects are governed by two
dimensionless scales, the first set by the ratio $\kappa=q/2M$
which characterizes the single--nucleon matrix
elements and the second by the ratio $\eta_F=p_F/M$ which enters whenever
the ground--state deuteron wave function occurs as in the matrix elements
making up the
deuteron break--up amplitude. Our objective here has been to explore the
degree to which quasifree electrodisintegration at high $q$ and $\omega$
decouples into {\bf relativistic} effects where expansions in
terms of $\kappa$ cannot be undertaken and {\bf exchange} effects whose
importance is governed by $\eta_F/\kappa$.

As far as the interference (exchange) terms are concerned, our results can be
summarized as follows: These terms are proportional to the
momentum--space integral of the overlap between deuteron wave functions
centered at momenta corresponding to the kinematics for the two outgoing
nucleons. Since these momenta differ significantly for large momentum transfer
values, the overlap is small and consequently the
interference effects are also small. Moreover,
the leading nonrelativistic contribution to the interference effects is
proportional to $G_E^p G_E^n$ for the longitudinal response and
${1\over 3}G_M^p G_M^n$ for the transverse response. Thus, when the overlap
is significant (low $q$), the electric neutron form factor is negligible,
and the longitudinal contribution to the interference terms is very small.
On the other hand, the transverse contributions are suppressed as well,
however now because of an extra
factor of $1/3$ of purely geometrical ({\it i.e.,\/} Clebsch-Gordan) origin.
Therefore, at least insofar as the nonrelativistic analysis
is concerned, it appears that the interference contributions to inclusive
deuteron electrodisintegration are usually relatively mild, except for
backward--angle
scattering and near threshold. These simple nonrelativistic arguments
appear also to be borne out in the present relativistic PWBA framework.

As far as the relativistic effects themselves are concerned, our analysis
which retains terms of all orders in $\kappa$ yields sizeable differences
when compared with a nonrelativistic expansion, as expected, especially
when kinematics beyond the quasifree region are explored. This is true
in the relativistic PWIA and continues to be true in the relativistic PWBA.

Thus, we see that indeed the
interference and relativistic effects largely decouple, especially in the
longitudinal response, the former usually being important only when
the momentum transfer is small, that is, for $\kappa<$ few$\times\eta_F$
corresponding to $q<$ few$\times (2p_F)$. Since the latter become
important when $\kappa\sim 1$, the effects usually decouple as long as
$p_F<\!< M$ ($\eta_F<\!< 1$), which is certainly true for the deuteron
where $p_F\cong$ 55 MeV/c (in contrast to heavier nuclei where it is
larger --- see below). There are exceptions to these statements, however, and
we have identified two cases
where the interference effects are important even at high momentum transfers.
The first is parity--violating electron scattering where,
due to the interesting
coincidence that $1-4\sin^2\theta_W\approx 0$, the interference effects
for the longitudinal PV response are enhanced by an order--of--magnitude.
A second is the quasi--deuteron model, where the wave function
overlap is significant even at high $q$ due to the larger Fermi momentum
for nuclei in general compared with deuterium which is exceptional
(for example, $\eta_F(^{40}$Ca$)/\eta_F(^{2}$H$)\cong 4.5$).
Effectively then the conditions $\kappa\sim$ few$\times\eta_F$ and
$\kappa\sim 1$ can be met simultaneously. The largeness of the overlap
also stems in part from the spin--0 nature of the quasi--deuteron
which eliminates the
factor of $1/3$ in $R_T$ occurring in the spin--1 deuteron case, thus leading
to an important transverse interference contribution. These observations
derived from our exploratory study of the relativistic PWBA within the
context of a simple quasi--deuteron model suggest further investigations
of nuclei with $A>2$. It is our intent to return in future work to study
both exceptions in more detail.

\newpage

\def\sf1{\sin\Phi}
\def\cf1{\cos\Phi}
\def\shf1{\sinh\Phi}
\def\chf1{\cosh\Phi}

\def\sy1{\sin\Psi}
\def\cy1{\cos\Psi}
\def\shy1{\sinh\Psi}
\def\chy1{\cosh\Psi}

\def\gm1{ G_M^{\tau_N} }
\def\ge1{ G_E^{\tau_N} }
\def\rt1{{\sqrt\tau}}
\def\pmb#1{\setbox0=\hbox{$#1$}%
\kern-.025em\copy0\kern-\wd0
\kern.05em\copy0\kern-\wd0
\kern-.025em\raise.0433em\box0 }
\def\ka1{ {\pmb\kappa} }
\def\hte{ {\pmb\eta}   }
\def\la1{\bar\lambda}
\def\ve1{\varepsilon}
%%%%%%%%%%%%%%%%%%%%%%%%%%%%%%%%%%%%%%%%

\section*{Appendix}
Let us first discuss Eq.~(\ref{torino_1}). We introduce the
dimensionless variables ${\pmb \kappa} = {\pmb q}/2M$, ${\pmb\eta}
={\pmb p}/M$, $\la1={\bar\w}/2M$,
$\varepsilon=\sqrt{{\pmb p}^2 + M^2}/M $
and $\tau=(q^2-{\bar\w}^2)/4M^2$, where we define
$\bar\w = E_N - E_B = \w + M_d - 2 M \varepsilon$ (the energy transferred
to an on--shell nucleon with momentum {\bf p}).
We also introduce two angles $\Psi$ and $\Phi$:
\begin{eqnarray}\label{angles}
\tan\Psi &\equiv &{ |\ka1\times\hte|\over 1+\tau+\ve1+\la1}\nonumber\\
\tanh\Phi &\equiv &{ |\ka1\times\hte|\over\sqrt{\tau}(\ve1+\la1)}\ .
\end{eqnarray}
Then for the charge operator we obtain from Ref.~\cite{Alberico}
\begin{eqnarray}\label{charge}
J_n^{(0)}& =&{1\over \sqrt{\ve1(\ve1+2\la1)]}} {\kappa\over\rt1}
\biggl[\ge1 \cosh\Phi\cos\Psi +\rt1 \gm1\sinh\Phi\sin\Psi\biggr]
\nonumber\\
J_s^{(0)}& =&{-1\over \sqrt{\ve1(\ve1+2\la1)}} {\kappa\over\rt1}
\biggl[ -G_E^{\tau_N} \cosh\Phi\sin\Psi +\rt1\gm1\sinh\Phi\cos\Psi\biggr]\ .
\end{eqnarray}
For the transverse projections we have
\begin{eqnarray}\label{curr}
J_n^{(1)} & = & {1\over \sqrt{\ve1(\ve1+2\la1)}}
\biggl[ \ge1\shf1\cy1 + \rt1\gm1\chf1\sy1\biggr]\nonumber\\
J_s^{(1)} & = & {-1\over \sqrt{\ve1(\ve1+2\la1)}}
\biggl[- \ge1\shf1\sy1 + \rt1\gm1\chf1\cy1\biggr]\nonumber\\
J_n^{(2)} & = & {-i\over \sqrt{\ve1(\ve1+2\la1)} }\gm1
\sqrt{(\rt1\chf1\sy1+\shf1\cy1)^2-(1+\tau)\sinh^2\Phi}\nonumber\\
J_s^{(2)} & = & {-i\over \sqrt{\ve1(\ve1+2\la1)} }\gm1
\biggl[\rt1\chf1\cy1 - \shf1\sy1\biggr]\ .
\end{eqnarray}
The longitudinal ($\mu=3$) projection of the 3--current is
related to the charge by the continuity equation
\begin{equation}\label{cont}
\kappa J^{(3)}_i =\ka1\cdot{\pmb J_i}=\lambda J^{(0)}_i \ .
\end{equation}
The $H$--type matrix elements in Eq.~(\ref{torino_1}) are obtained by
replacing ${\pmb p_B}$ with ${-\pmb p_N}$, $\tau_N$ with $\tau_B$
 and $\lambda$ with $-\lambda$. It is straightforward
to check that this results in just reversing the sign of
$\sin\Psi$ and $\sinh\Phi$.
The extreme nonrelativistic limit (that is, to ${\cal O}[1/M]^0$)
of Eqs.~(\ref{angles}--\ref{curr})
is presented in Table~II, whereas a higher--order nonrelativistic
reduction can be found in Ref.~\cite{Alberico}.

We now turn to Eq.~(\ref{exchange_sum}).
The coefficients $a^{K,X}_{i,j}$ are defined as follows:
\begin{eqnarray}\label{coefficients}
a^{K,X}_{i,j}& = &\sum_{M_S,M_S'}\sum_{m,m'}
\sum_{m_N,m_B} (-1)^{1+M_S}\three_j(1,1,K,M_S,-M_S',X)\times\nonumber\\
& &\three_j(\half,\shalf,1,m_N,m',-M_S')\three_j(\shalf,\shalf,1,m_B,m,-M_S)
\theta_{i,j} \; g^{\mu}_{m'}\; g^{\nu}_{m} \  ,
\end{eqnarray}
where we have set
\begin{eqnarray}\label{lalala}
\theta_{n,n} &\equiv & \delta_{m_B,m'}\delta_{m,m_N}\\
\theta_{n,s} &\equiv & 2 m \delta_{m',m_B}(1-\delta_{m,m_N})
                \nonumber\\
\theta_{s,n} &\equiv & 2 m' \delta_{m,m_N}(1-\delta_{m',m_B})
                \nonumber\\
\theta_{s,s} &\equiv & 4 m m' (1-\delta_{m,m_N}-\delta_{m',m_B}
        +\delta_{m',m_B}\delta_{m,m_N})\nonumber \ ,
\end{eqnarray}
with
\begin{equation}\label{hat}
g^{\mu}_{m}=\cases{1   &, if $\mu \neq 2$\cr
                   2 m  &, if $\mu =2.$   \cr}
\end{equation}
No summation convention over repeated indices is implied here.
In Table~I we write down explicitly the coefficients $a^{K,X}_{i,j}$.
We use a block--diagonal matrix notation in order to bring out the
different behaviour of the (2)--components as seen in Eq.~(\ref{3_currents}).

Finally, let us calculate the longitudinal and transverse responses
in the S--state--only limit using Eqs.~(\ref{direct_sum}--\ref{exchange_sum})
and Table~I. For the longitudinal response we obtain
\begin{eqnarray}\label{longitudinal_s_only}
R_L^{S-state} &\sim&
\Bigl[ |J_{(n)}^0|^2 +  |J_{(s)}^0|^2\Bigr] |u(p)|^2 +
\Bigl[ |H_{(n)}^0|^2 +  |H_{(s)}^0|^2\Bigr] |u(p_N)|^2 \nonumber\\
& &+ 2u(p)u(p_N)\Re\Bigl[ \bigl\{H_{(n)}^{*0}J_{(n)}^0\bigr\}
+\left(1\over 3\right )  \bigl\{H_{(s)}^{*0}J_{(s)}^0\bigr\}\Bigr] \ .
\end{eqnarray}
Similarly, for the transverse response we have
\begin{eqnarray}\label{transverse_s_only}
R_T^{S-state} &\sim &
\Bigl[ |J_{(n)}^1|^2 +  |J_{(s)}^1|^2 +
 |J_{(n)}^2|^2 +  |J_{(s)}^2|^2\Bigr] |u(p)|^2 \nonumber\\
& &+\Bigl[ |H_{(n)}^1|^2 +  |H_{(s)}^1|^2 +
 |H_{(n)}^2|^2 +  |H_{(s)}^2|^2\Bigr] |u(p_N)|^2 \nonumber\\
& &+ 2u(p)u(p_N)\Re\biggl[ \bigl\{H_{(n)}^{*1}J_{(n)}^1\bigr\}
+\left(1\over 3\right) \bigl\{H_{(n)}^{*2}J_{(n)}^2\bigr\}\nonumber\\
& &+\left(1\over 3\right )  \bigl\{H_{(s)}^{*1}J_{(s)}^1\bigr\}
+\left(1\over 3\right )\bigl\{H_{(s)}^{*2}J_{(s)}^2\bigr\}
\biggr] \ .
\end{eqnarray}

\newpage
%definitions
\def\PL{ {\it Phys. Lett.} }
\def\NP{ {\it Nucl. Phys.} }
\def\PTP{ {\it Prog. Theor. Phys.} }
\def\PRL{ {\it Phys. Rev. Lett.} }
\def\PR{ {\it Phys. Rev.} }

\newpage
$$\hbox{\vbox{\offinterlineskip
\def\strut{\hbox{\vrule height 15pt depth 10pt width 0pt}}
\hrule
\halign{
\strut\vrule#\tabskip 0.05in&
\hfil$#$\hfil&
\vrule#&
\hfil$#$\hfil&
\vrule#&
\hfil$#$\hfil&
\vrule#&
\hfil$#$\hfil&
\vrule#&
\hfil$#$\hfil&
\vrule#\tabskip 0.0in\cr
&\multispan{9}\hfil{\bf Table I\quad The $a^{K,X}_{i,j}$
coefficients  $\left(\matrix{\mu,\nu\neq 2& \nu=2\cr
                             \mu=2 & \mu=\nu=2\cr}\right) $  }\hfil&
        \cr\noalign{\hrule}
& (K,X)  && NN && NS && SN && SS  &
\cr\noalign{\hrule}
& (0,0) &&{1\over \sqrt{3}}\xara(1,0,0,{1\over 3}) && 0&& 0 &&
        {1\over 3\sqrt{3}} \xara(1,0,0,1)
                 & \cr\noalign{\hrule}
%& && && && && & \cr\noalign{\vskip - .7 true cm\hrule}
&\multispan{9}\hfil&
        \cr\noalign{\vskip -.2 true cm\hrule}
& (1,-1)&& 0 && {1\over 3\sqrt{3}} \xara(1,1,0,0)&&
         {1\over 3\sqrt{3}} \xara(-1,0,1,0)&& 0 & \cr\noalign{\hrule}
& (1,0 )&& {\sqrt{2}\over 3\sqrt{3}}\xara(0,1,1,0) &&   0   &&  0 && 0 &
 \cr\noalign{\hrule}
& (1,+1)&& 0 && {1\over 3\sqrt{3}} \xara(1,-1,0,0)&&
         {1\over 3\sqrt{3}} \xara(-1,0,-1,0)&& 0 & \cr\noalign{\hrule}
%&       &&   &&               &&             &&   & \cr\noalign{\hrule}
&\multispan{9}\hfil&
        \cr\noalign{\vskip -.2 true cm\hrule}
& (2,-2)&& 0 &&      0        &&     0       && {
1\over 3\sqrt{5}}\xara(-1,-1,1,1)
                         & \cr\noalign{\hrule}
& (2,-1)&& 0 &&       {1\over 3\sqrt{5}} \xara(0,0,1,1)&&
 {1\over 3\sqrt{5}} \xara(0,-1,0,1)  && 0 &   \cr\noalign{\hrule}
& (2,0 )&&{2\sqrt{2}\over 3\sqrt{15}}\xara(0,0,0,1) &&      0
 &&     0       && -{\sqrt{2}\over 3\sqrt{15}}\xara(1,0,0,1) &
                                                  \cr\noalign{\hrule}
& (2,+1)&& 0 &&   {1\over 3\sqrt{5}}\xara(0,0,1,-1)   &&  {1\over 3\sqrt{5}}
\xara(0,-1,0,-1)       && 0 & \cr\noalign{\hrule}
& (2,+2)&& 0 &&      0        &&     0       && {1\over 3\sqrt{5}}
\xara(-1,1,-1,1)  &   \cr\noalign{\hrule}
}}}$$
\bigskip
$$\hbox{\vbox{\offinterlineskip
\def\strut{\hbox{\vrule height 15pt depth 10pt width 0pt}}
\hrule
\halign{
\strut\vrule#\tabskip 0.3in&
\hfil$#$\hfil&
\vrule#&
\hfil$#$\hfil&
\vrule#&
\hfil$#$\hfil&
\vrule#&
\hfil$#$\hfil&
\vrule#&
\hfil$#$\hfil&
\vrule#\tabskip 0.0in\cr
&\multispan{9}\hfil{\bf Table II\quad Extreme Nonrelativistic Limit of
Eq.~(\ref{torino_1})}\hfil&\cr\noalign{\hrule}
& (\mu)&& J^{\mu}_{(n)} && H^{\mu}_{(n)} &&  J^{\mu}_{(s)}&&
H^{\mu}_{(s)}& \cr\noalign{\hrule}
& (0) && G_E^{\tau_N} &&  G_E^{\tau_B}&& 0  && 0& \cr\noalign{\hrule}
& (1) && 0  && 0 && -\kappa\,G_M^{\tau_N} &&
-\kappa\,G_M^{\tau_B}&\cr\noalign{\hrule}
& (2) && 0  && 0  &&i \kappa\,G_M^{\tau_N}
&&i\kappa\,G_M^{\tau_B }&\cr\noalign{\hrule}
}}}$$

\newpage
$$\hbox{\vbox{\offinterlineskip
\def\strut{\hbox{\vrule height 15pt depth 10pt width 0pt}}
\hrule
\halign{
\strut\vrule#\tabskip 0.3in&
\hfil$#$\hfil&
\vrule#&
\hfil$#$\hfil&
\vrule#&
\hfil$#$\hfil&
\vrule#&
\hfil$#$\hfil&
\vrule#\tabskip 0.0in\cr
&\multispan{7}\hfil{\bf Table III.1\quad The ${\cal D}$ and ${\cal E}$
integrals at $q=150$ MeV/c}\hfil&\cr\noalign{\hrule}
& \w\,\hbox{(MeV)}  && {\cal D} &&{\cal E} &&{\cal E/D}  & \cr\noalign{\hrule}
& 8.3 && 0.939 && 0.912 && 0.971 & \cr\noalign{\hrule}
& 10  && 5.076 && 3.064 && 0.604 & \cr\noalign{\hrule}
& 16  && 8.930 && 2.563 && 0.287 & \cr\noalign{\hrule}
& 26  && 2.943 && 0.975 && 0.331 & \cr\noalign{\hrule}
& 36  && 1.107 && 0.429 && 0.388 & \cr\noalign{\hrule}
}}}$$
\bigskip
$$\hbox{\vbox{\offinterlineskip
\def\strut{\hbox{\vrule height 15pt depth 10pt width 0pt}}
\hrule
\halign{
\strut\vrule#\tabskip 0.3in&
\hfil$#$\hfil&
\vrule#&
\hfil$#$\hfil&
\vrule#&
\hfil$#$\hfil&
\vrule#&
\hfil$#$\hfil&
\vrule#\tabskip 0.0in\cr
&\multispan{7}\hfil{\bf Table III.2\quad The ${\cal D}$ and ${\cal E}$
integrals at $q=300$ MeV/c}\hfil&\cr\noalign{\hrule}
& \w\,(\hbox{MeV})  && {\cal D} &&{\cal E} &&{\cal E/D} &\cr\noalign{\hrule}
& 27  && 0.100 && 0.079 && 0.784 & \cr\noalign{\hrule}
& 30  && 0.285 && 0.136 && 0.475 & \cr\noalign{\hrule}
& 40  && 2.835 && 0.301 && 0.106 & \cr\noalign{\hrule}
& 50  && 4.846 && 0.208 && 0.043 & \cr\noalign{\hrule}
& 100 && 0.222 && 0.003 && 0.014 & \cr\noalign{\hrule}
}}}$$

%%%%%%%%%%%%%%%%%%%%%%%%%%%%%%%%%%%%%%%%%%%%%%%%%%%%%%%%%%%%%%%%%%%%%%%%%%

%  FIGURE CAPTIONS

\clearpage\newpage
\centerline{\bf Figure Captions}
\begin{enumerate}
\item[1.] Kinematics for single--arm coincidence electron scattering.
          Besides the outgoing electron $K^{\mu}$, the nucleon labelled
          $P^{\mu}_N$ is also detected.
\item[2.] Electrodisintegration of deuterium in the PWBA,
          with nucleon $N$ detected in coincidence.\hfill\break
          (a): Amplitude corresponding to the PWIA where the detected  nucleon
             ($N$) is the one that reacted with the virtual photon.\hfill\break
          (b): While the detected nucleon is  still nucleon ($N$), it is
               nevertheless  nucleon $B$ that has reacted. This amplitude is
               ignored in the PWIA.
\item[3.] The hadronic tensor for inclusive deuteron electrodisintegration in
          the PWBA.
          The momenta and spins of the nucleons involved are indicated.
\item[4.] The Direct (${\cal D}$) and exchange (${\cal E}$) integrals at
	  $q=150$ MeV/c:
          (a) near threshold ($\w=9$ MeV) and (b) on the quasielastic peak
          ($\w=14.13$ MeV).
\item[5.] The ratio ${\cal E}/{\cal D}$ as a function of momentum transfer
          $\w$ for (a) $q=150$ MeV/c and (b) $q=300$ MeV/c. The arrows
          mark the position of the quasielastic peak in each case.
\item[6.] The ${\cal E}/{\cal D}$ ratio as a function of the momentum
          transfer $q$ for energy transfer fixed to the quasielastic
          peak (Eq.~(40)).
\item[7.] Forward angle ($\theta_e=35^o$) parity--violating asymmetry
          in the PWBA (solid lines) and PWIA (dashed lines) as a function
          of the energy transfer $\omega$ for fixed momentum transfer:
          (a) $q=150$ MeV/c and (b) $q=300$ MeV/c. The position of the
          quasielastic peak is denoted by an arrow in each case.
\item[8.] Percentage contours for the absolute value of the ratio
          $\cal{ E}/{\cal D}$ in the $q-\w$ plane. The shaded area corresponds
          to quasifree kinematics calculated according to the Fermi--Gas
          formula with $p_F=60$ MeV/c (see text).
\item[9.] Percentage contours for the difference between PWIA
          and PWBA as represented by the ratio $\Delta_{BF/IF}$ (Eq.~(43)) for
          the longitudinal response $R_L$ in inclusive electrodisintegration
          of deuterium.
\item[10.] Same as in Fig.~9, except now for the transverse response $R_T$.
\item[11a.] Contour plot for the ratio $\Delta_{BSF/BN}$ (Eq.~(44))
             in the case of the longitudinal response.
\item[11b.] Same as in Fig.~11a, but with the PWBA modified by $(1+\tau)$
           (see text).
\item[12a.] Same as in Fig.~11a, except now for the transverse response $R_T$.
\item[12b.] Same as in Fig.~12a, but with the PWBA modified by $\tau/\kappa ^2$
           (see text).
\item[13.]  Same as in Fig.~8, but with artificial quasi--deuteron $J=0$
           wave functions
           corresponding to $p_F=300$ MeV/c and binding energy $E_b=16$ MeV.
\item[14.] Same as in Fig.~9, except now for the quasi--deuteron.
\item[15.] Same as in Fig.~10, except now for the quasi--deuteron.
\end{enumerate}
\qqq